\documentclass[prx,twocolumn,a4paper,showkeys,showpacs]{revtex4-1}
\usepackage[dvipdfmx]{graphicx}
\usepackage{bmpsize}
\usepackage{calc}
\usepackage{amsmath}
\usepackage{physics}
\usepackage{bm}         
\usepackage{mathrsfs}  
\usepackage{amssymb}
\usepackage{comment}
\usepackage{cases}
\usepackage{here}
\usepackage{array}
\usepackage{xr}
\usepackage{color}
\usepackage{tabularx}
\newcolumntype{C}{>{\centering\arraybackslash}X}
\newcolumntype{L}{>{\raggedright\arraybackslash}X}
\newcolumntype{R}{>{\raggedleft\arraybackslash}X}
\usepackage{ulem}

\usepackage{lipsum}

\begin{document}

\title{
Selective decision making and collective behavior of fish by the motion of visual attention
}
\author{Susumu Ito and Nariya Uchida}
\thanks{uchida@cmpt.phys.tohoku.ac.jp}
\affiliation{Department of Physics, Tohoku University, Sendai, 980-8578, Japan}
\date{\today} 

\begin{abstract}
Collective motion provides a spectacular example of self-organization in Nature.
Visual information plays a crucial role among various types of information in determining interactions.
Recently, experiments have revealed that organisms such as fish and insects
selectively utilize a portion, rather than the entirety, of visual information.
Here, focusing on fish,
we propose an agent-based model where the direction of attention is guided by visual stimuli
received from the images of nearby fish.
Our model reproduces a branching phenomenon
where a fish selectively follows a specific individual as the distance between two or three nearby fish increases.
Furthermore, our model replicates various patterns of collective motion in a group of agents,
such as vortex, polarized school, swarm, and turning.
We also discuss the topological nature of visual interaction,
as well as the positional distribution of nearby fish and
the map of pairwise and three-body interactions induced by them.
Through a comprehensive comparison with existing experimental results,
we clarify the roles of visual interactions and issues to be resolved by other forms of interactions.
\end{abstract}

\maketitle

\section{Introduction}
Collective motion and cluster formation are ubiquitously found in various organisms~\cite{Vicsek2012}.
Fish are no exception: there are swarm in which the direction of motion of fish is random,
polarized school which shows directional movement,
and vortex in which fish rotates around an axis~\cite{Parrish2002,Lopez2012,Tunstrom2013,Terayama2015}.
Schooling is induced not only by hydrodynamic interactions~\cite{Partridge1980,Filella2018,Li2020,Ito2023},
but also for efficiently foraging food~\cite{Harpaz2020}
and avoiding predator~\cite{Parrish2002,Radakov1973},
in which visual cues play an important role
in transferring information among individuals~\cite{Harpaz2020,Peshkin2013,Rosenthal2015,Poel2022,Hein2022}.

How does a fish process visual information in a school?
Reacting to all neighboring fish within the field of view imposes a heavy load 
on the information processing system~\cite{Dukas1998},
and it is considered to be less suitable for fish with relatively small brains~\cite{Herbert2011,Hein2022}.
In fact, zebrafish (\textit{Danio rerio})
show selective decision making~\cite{Burgess2010,Sridhar2021}:
when there are two or three targets (i.e. light spots or virtual fish)
within a certain distance in the range of eye sight,
a zebrafish is attracted to one of the targets.
In other words, a zebrafish selects and pays attention to one target, by truncating the visual information from other targets.
A similar behavior is observed for fly and locust~\cite{Sridhar2021},
indicating the commonness of selective decision making through visual information.

Fish schools have been modeled by agent-based models
for decades~\cite{Breder1954,Aoki1982,Huth1992,Huth1994,Niwa1994}
with simple geometric criteria to determine the interaction:
metric range~\cite{Couzin2002,Kunz2003,Hemelrijk2008},
topological distance~\cite{Inada2002,Viscido2005,Ito2022a,Ito2022b},
and Voronoi tessellation~\cite{Gautrais2012,Calovi2014,Calovi2015,Filella2018,Deng2021,Liu2023}.
Recent studies incorporate visual information for modeling the collective motion of fish~\cite{Kunz2012},
bird~\cite{Pearce2014},
and generic agents~\cite{Lemasson2009,Lemasson2013,Bastien2020,Qi2023,Castro2023}.
These models assume that an agent can integrate
the information (e.g. distance and velocity) of all neighbors detected
to decide its action.
Some models~\cite{Collignon2016,Sridhar2021,Gorbonos2023,Oscar2023}
treat selective decision making under specific conditions.
Collignon \textit{et al.}~\cite{Collignon2016} used a model
stochastic model to reproduce the motion of 10 fish exhibiting burst-coast swimming
in a tank with feeders.
Sridhar \textit{et al.}~\cite{Sridhar2021} and the successive studies~\cite{Gorbonos2023,Oscar2023}
introduced a network model inspired by the Ising spin model and its phase transition.
This model was applied to an agent interacting with a few virtual agents,
and explained a bifurcation behavior of the agent's position with the change of the distance between the targets.

In this paper, a comprehensive model of selective decision making 
and resultant collective motion based on visual stimuli.
We introduce the notion of direction of attention induced by the stimulus
and model its movement induced by the visual stimuli from neighbor fish.
The fish interact via a repulsive, attractive and alignment interactions 
with the neighbors in the direction of attention.
For a few fish, our model reproduces the bifurcation behavior observed in the previous experiment~\cite{Sridhar2021}.
For many fish, our model exhibits various patterns of emergent collective motion
(vortex, polarized school, swarm and turning).
Furthermore, we analyze the topological distance~\cite{Herbert2011}
and the pairwise and three-body forces~\cite{Katz2011}
from the viewpoint of visual interaction.
The roles of visual information in the movement of fish is
discussed by comprehensive comparison with the experimental results.

\begin{figure*}[!t]
\centering
\includegraphics[width=\linewidth]{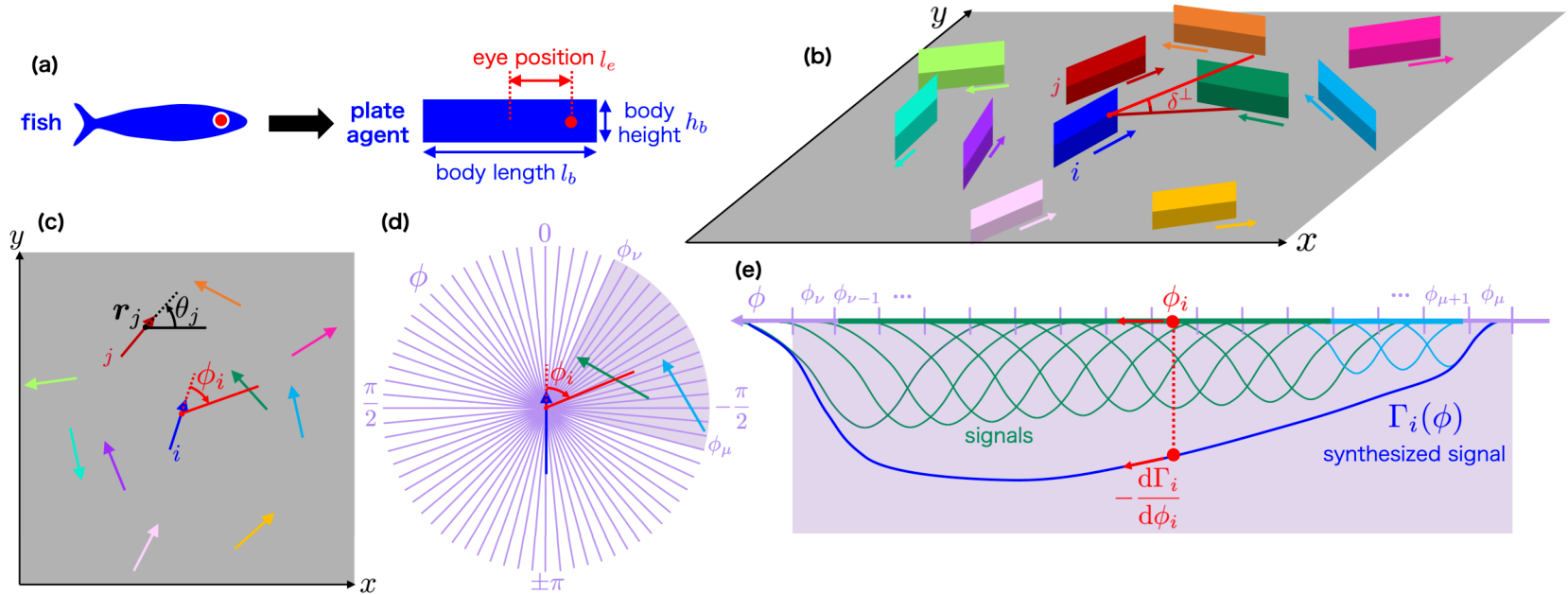}
\caption{
Schematic illustration of our model.
(a) An agent is a rectangular plate of length $l_b$ and height $h_b$.
The eye is positioned at distance $l_e$ from the body center.
(b) The agents move in a $x$-$y$ plane. Each arrow represents the direction of motion, and
$\delta^\perp$ is the vertical angular diameter
of the green agent measured by the eye of the blue agent ($i$). It varies over the body of the green agent.
(c) The top view of (b). The angle $\phi_i$ gives the direction of visual attention of $i$ relative to the head direction.
$\bm{r}_j=(x_j,y_j)$ and $\theta_j$ are the eye position and the angle of orientation of $j$.
Both of the angles are measured in the couterclockwise direction and take values in $[-\pi,\pi]$.
(d) The visual field of the blue agent (i) is a circle parametrized by the angle $\phi \in [-\pi, \pi]$ and separated into bins. The two neighbors (green and cyan) are detected in the bins in  $\phi_\mu \le \phi \le \phi_\nu$.
(e) The signals from the images in $[\phi_\mu,\phi_\nu]$ in (d).
The green and cyan lines on the $\phi$ axis show the bins in the visual field
occupied by each agent.
The green and cyan curves show the signals $\gamma_{i,\lambda}(\phi)$ from the $\lambda$-th bin
that are superposed as the synthesized signal $\Gamma_i(\phi)$ in the perception field.
The red arrow shows the derivative -$\dv*{\Gamma_i}{\phi_i}$.
}
\label{schematic}
\end{figure*}

\section{Model}
Our model is based on the eye system and experimentally observed behaviors of fish.
First, neighbor fish are detected as projected images on the retina,
and the signal is transmitted to the visual centers of the brain
through retinal ganglion cells~\cite{Pita2015}.
The resolution of the images are determined by the number of ganglion cells. 
Their density is on the order of 10$^4$ cells/mm$^2$
for zebrafish and golden shiner (\textit{Notemigonus crysoleucas}),
which use visual information as primary source of information.
In addition, the resolution is higher in the forward direction than the backward:
the density of the ganglion cells is higher in the rear side of the retina than in the front~\cite{Pita2015}.

Second, an experiment on larval zebrafish shows that
the visual interaction mainly depends on the vertical size of the image,
and starts to depend weakly on the horizontal size as they grow~\cite{Harpaz2021b}.
It is considered that fish adopts the vertical size as visual information
to precisely measure the distance to neighbors,
because the horizontal size easily changes by the relative heading angle of the neighbor
as the fish has a slender body~\cite{Harpaz2021b}.

Third, the eyeball of fish moves well~\cite{Kabayama1979},
and fish can track moving targets~\cite{Kawamura1978}.
The motion velocity of an image as well as its size is encoded on the brain~\cite{Hein2022}.
An experiment using golden shiner shows that
a fish tends to follow the neighbors at a close distance in the front,
as well as those moving at high relative speed.~\cite{Sridhar2023}.
Therefore, we incorporate the relative position and speed of the neighbor
agents as the key elements for visual interaction.

Finally, the interaction forces are classified into repulsion, attraction, and alignment.
Repulsion at short distances and attraction at large distances are demonstrated by a force map
for golden shiner~\cite{Katz2011} and mosquitofish (\textit{Gambusia holbrooki})~\cite{Herbert2011}.
The aligning interaction is found later
for rummy-nose tetra (\textit{Hemigrammus rhodostomus})~\cite{Calovi2018}.
It is also worth mentioning that the orientational order of zebrafish
is not explained by repulsion and attraction only~\cite{Harpaz2021b}.
Integrating these properties, we construct our model as follows.

We model each self-propelled agent (fish) as a rectangular plate of length $l_b$, height $h_b$
which has a monoeye at distance $l_e$ its center (Fig.~\ref{schematic}(a)),
and consider $N$ agents moving on a plane with no boundary (Fig.~\ref{schematic}(b)).
For the agent indexed by $i=1,2,\cdots,N$, we define the position of the eye $\bm{r_i}=(x_i,y_i)$ and
the velocity $\bm{v}_i=\dv*{\bm{r}_i}{t}=v_i\bm{e}_i$, where
$v_i\geq0$ is the speed, and $\bm{e}_i=(\cos\theta_i,\sin\theta_i)$ as the unit vector in the direction of motion
($\theta_i$ is the angle of orientation).

The essential new feature of our model is the direction of visual attention
specified by its angle $\phi_i$ (see Fig.~\ref{schematic}(c)).
We model the motion of $\phi_i$ as follows.
As shown in Fig.~\ref{schematic}(d),
we divide the visual field into bins,
and the number of bins $N_b$ is determined by the density of the ganglion cells.
The angle of the $\mu$-th bin ($\mu=1,2,\cdots,N_b$) is denoted by $\phi_\mu$.
The vertical angular diameter $\delta_{i,\mu}^\perp$ (see Fig.~\ref{schematic}(b))
and the relative speed $u_{i,\mu}$ of the neighbor in each bin is encoded as the visual information.

The image in each bin is transmitted as a signal that has a distribution in the ``perception field'' $\phi \in [-\pi, \pi]$, which is described by a smooth signal function $\gamma_{i,\mu}(\phi)$:
\begin{equation}
\label{eq gamma}
\gamma_{i,\mu}(\phi)=-U(\delta_{i,\mu}^\perp,u_{i,\mu};\beta)A(\delta_{i,\mu}^\perp)D(\phi_{\mu};\chi)G(\phi,\phi_{\mu};\kappa).
\end{equation}
The angle dependence is given by
the periodic function of $\phi$, 
\begin{equation}
\label{eq G}
G(\phi,\phi_{\mu};\kappa)=\exp[\kappa\{\cos(\phi-\phi_\mu)-1\}],
\end{equation}
where $\kappa>0$ gives the sharpness of a signal. 
Note that $G=1$ at $\phi=\phi_\mu$, where the image is located.
The function $D(\phi_{\mu};\chi)$ represents the density of ganglion cells 
and has the aforementioned front-back asymmetry parametrized by $\chi$.
The dependences on the vertical angular diameter 
and the relative speed of the image are described by
$A(\delta_{i,\mu}^\perp)$ and
$U(\delta_{i,\mu}^\perp,u_{i,\mu};\beta)$, respectively.
(See \textit{Materials and Methods} for details.)

The signals are superposed as a synthesized signal (see Fig.~\ref{schematic}(e)):
\begin{equation}
\label{eq Gamma}
\Gamma_i(\phi)=\sum_{\mu=1}^{N_b}\gamma_{i,\mu}(\phi),
\end{equation}
and the angle of direction of attention $\phi_i$ is driven by the equation of motion
\begin{equation}
\label{eq phi}
\tau_\phi\dv{\phi_i}{t}=-\dv{\Gamma_i(\phi_i)}{\phi_i},
\end{equation}
where $\tau_\phi$ is the characteristic time scale for $\phi_i$.
In other words, $\phi_i$
tends to approach a the local minimum of $\Gamma_i(\phi)$
which is regarded as a potential function in the perception field, and
the local minimum is deeper when the neighbor
is closer, and further forward, and/or has higher relative speed.

Next, we introduce the equation of motion of the position of the $i$-th agent.
We adopt the variables ($v_i,\theta_i$) for the equation of motion~\cite{Bastien2020,Qi2023}
because, in addition to the change of the orientation,
the change of the speed is also important for fish~\cite{Katz2011,Herbert2011}.
The equation for $v_i$ is
\begin{equation}
\label{eq v}
\dv{v_i}{t}=C(v_0^2-v_i^2)+\langle F(\phi_\mu,\delta^\perp_{i,\mu})\rangle_\mu+\eta_{v,i},
\end{equation}
and the equation for $\theta_i$ is
\begin{equation}
\label{eq tht}
\dv{\theta_i}{t}=\langle\Omega(\phi_\mu,\delta^\perp_{i,\mu},\psi_{i,\mu})\rangle_\mu+\eta_{\theta,i},
\end{equation}
where the mass is normalized as one, $v_0$ is the steady swimming speed, and $C$ is a constant.
The self-propelled term $C(v_0^2-v_i^2)$ corresponds to
the balance between Newton's drag force $-Cv_i^2$ and the thrust force $Cv_0^2$~\cite{Gazzola2014}.
The speeding force term $F(\phi_\mu,\delta_{i,\mu})$ 
includes the repulsion and attraction,
and the angular velocity term $\Omega(\phi_\mu,\delta_{i,\mu},\psi_{i,\mu})$
includes the alignment in addition.
$\psi_{i,\mu}$ is the relative heading angle between the focal agent and the neighbor detected at $\phi_\mu$.
$\langle\cdots\rangle_\mu$ is the average over the resolution angle region.
Moreover, $\eta_{v,i}\dd t=\sqrt{2D_v}\dd w_{v,i}$ and $\eta_{\theta,i}\dd t=\sqrt{2D_\theta}\dd w_{\theta,i}$
correspond to the white Gaussian noises where $\dd w_{v,i}$ and $\dd w_{\theta,i}$ are the standard Wiener processes.
In our model, we rescaled the parameter values for non-dimensionalization by
a body length (BL) $l_b $, timescale 1 sec, and body mass.
(For example, the case in which the speed takes 1 means that the speed is 1 BL/sec.)
The main control parameters of our model are the heterogeneity of the density of ganglion cells $\chi$ 
and the strength of alignment $\omega_o$.
(See \textit{Materials and Methods} 
and \textit{SI Text} 
for details of the formulation and definitions of the quantities measured in the simulation.)

\section{Results}

\begin{figure}[!t]
\centering
\includegraphics[width=\linewidth]{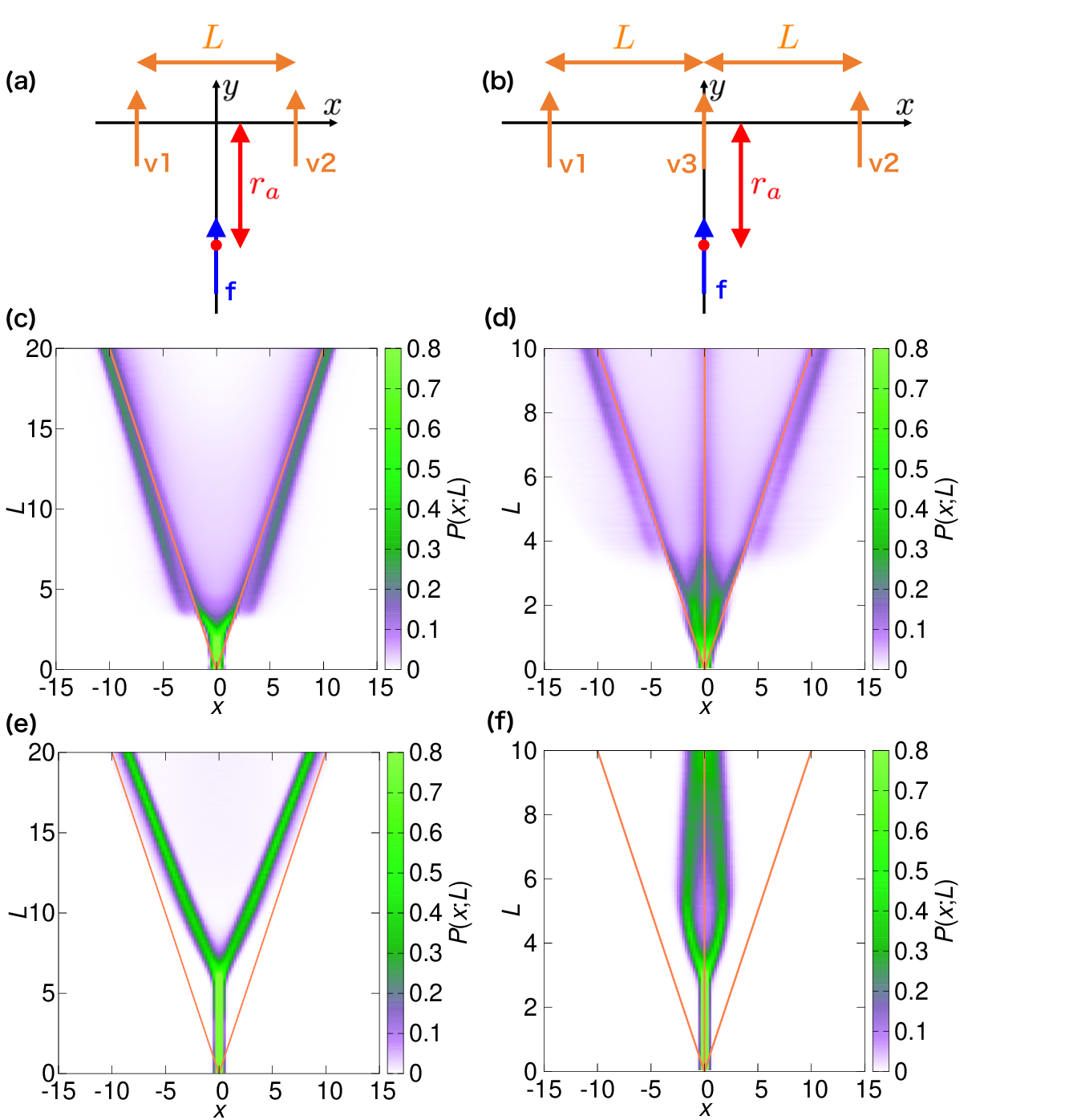}
\caption{
Selective decision making for (a),(c),(e) two and (b),(d),(f) three virtual agents with $\chi=0.3,\omega_o=1.0$.
(a)-(b) Schematic illustration of initial condition.
A Blue arrow is a focal agent f which will move freely in $x$-$y$ plane,
and the orange arrows are the virtual agents vn which will move along $y$-axis.
The marginal probability distribution $P(x;L)$ of (c)-(d) our model
and (e)-(f) the conventional particle model.
The orange lines represent the $x$ position of virtual agents.
}
\label{bifurcation}
\end{figure}

\begin{figure*}[!t]
\centering
\includegraphics[width=\linewidth]{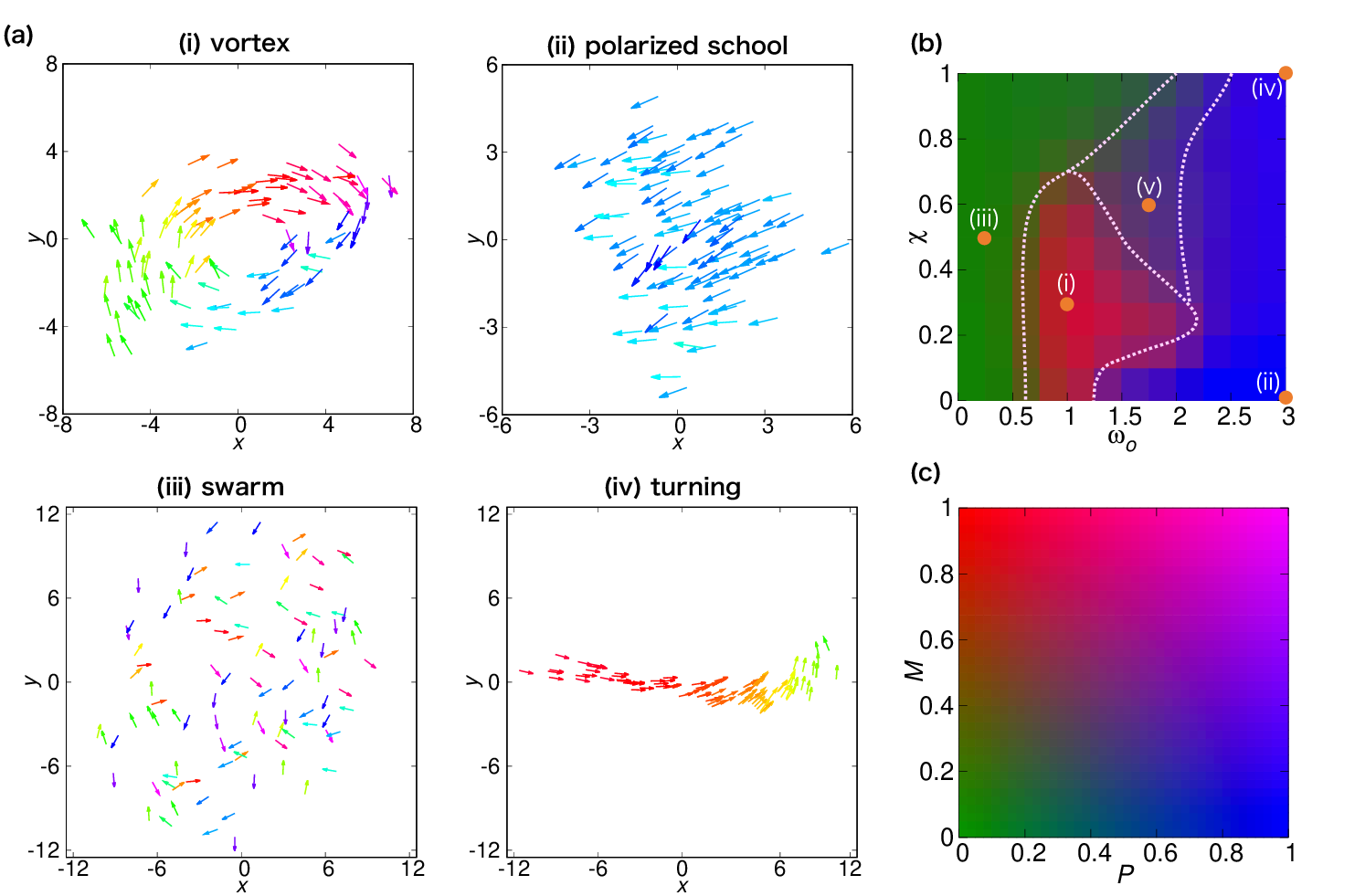}
\caption{
The snapshots of cluster and the phase diagram of pattern.
(i)-(v) correspond to the pattern and the parameters $(\chi,\omega_o)$ as follows.
(i) vortex: $\chi=0.3,\omega_o=1.0$.
(ii) polarized school: $\chi=0,\omega_o=3.0$.
(iii) swarm: $\chi=0.5,\omega_o=0.25$.
(iv) turning: $\chi=1.0,\omega_o=3.0$.
(v) unsteady aggregation: $\chi=0.6,\omega_o=1.75$.
(a) The snapshots of 100 agents which are represented as arrows of a body length $l_b$.
The color corresponds to the direction of motion.
(b) The phase diagram with respect to $(\chi,\omega_o)$.
The color represents the value of order parameters $(P,M)$ as shown in (c).
The orange points correspond to each parameter $(\chi,\omega_o)$ of (i)-(v).
The dashed lines represent the qualitative boundary of the patterns.
}
\label{snap pattern}
\end{figure*}

\subsection{Behavior of selective decision making}

First, we demonstrate that our model exhibits selective decision making by 
reproducing some key features of the experimental results in~\cite{Sridhar2021}.
Following the study, we consider a focal agent following a few ``virtual agents''
that perform prescribed motion without interacting with the other agents.
In the initial state ($t=0$), 
as shown in Fig.~\ref{bifurcation}(a)-(b), 
the focal agent f located at $x_\mathrm{f}=0$, $y_\mathrm{f}=-r_a$ 
is moving in the $y$-direction with the speed $v_\mathrm{f}=v_0$ 
and the angle of visual attention $\phi_\mathrm{f}=0$.
The virtual agents vn (n$=1,2, 3$ are located at $y_\mathrm{vn}=0$ with the lateral distance $L$ 
and move along the $y$-axis with the speed $v_\mathrm{vn}=v_0$.
The focal agent f obeys the time-evolution equations~(\ref{eq phi})-(\ref{eq tht})
and follows the virtual agents, which maintain their relative distance
and velocity in the $y$-direction.
Here we set the parameters as $\chi=0.3$ and $\omega_o=1.0$.

The focal agent fluctuates between different agents 
exhibiting a back-and-forth motion (Figs.~S3(a),(d)),
which has a characteristic timescale on the order of 10 time unit.
Trajectories plotted in the moving frame of the virtual agents 
show that the focal agent follows the virtual agents from the back
(Figs.~S3(b),(e)).

To study the $L$-dependence,
we use the positional distribution of the focal agent $P(x,y;L)$ 
in the moving frame 
and the marginal probability distribution $P(x;L)$ which is obtained by 
integrating $P(x,y;L)$ over the $y$-axis.
The plots of $P(x,y;L)$ in Figs.~S4,~S5 show that
the focal agent tends to be located behind the center of the virtual agents
for a small $L$,  
while the peak is split and located just behind each virtual agent
for larger values of $L$.
This bifurcation behavior is more clearly seen in the plots of $P(x;L)$ 
in Fig.~\ref{bifurcation}(c),(d),
which resemble the experimental results for fish~\cite{Sridhar2021}.
For two virtual agents (Fig.~\ref{bifurcation}(c)),
the focal agent is located at the center of the virtual agents for $L\lesssim3$,
and at the position of the virtual agents for $L\gtrsim3$.
For three virtual agents (Fig.~\ref{bifurcation}(d)),
$P(x;L)$ shows a three-way fork through two consecutive bifurcations;
for $L\lesssim1$, 
the focal agent is at the center of the three virtual agents; 
for $1\lesssim L\lesssim3$,
it resides at the center of v1 and v3 or
the center of v2 and v3; 
for $L\gtrsim3$,
it comes to the position of the virtual agents.
At a very large distance ($L\sim10$), 
the probability to follow the center virtual agent v3
becomes smaller than those for v1 and v2.
See \textit{SI Text} and Figs.~S6,S7
for dependence on the other parameters.

We also tested an asymmetrical configuration of three virtual agents,
where v3 is  shifted to the right
by the distance $L_{\mathrm{asym}}$ from the center
(see Fig.~S8(a)).
For  intermediate values of $L_{\mathrm{asym}}$,
the marginal probability distribution exhibits three peaks
at the start of the three-way fork bifurcation ($L=3.0$),
instead of the four peaks for $L_{\mathrm{asym}}=0$.
The peaks are located at the right of v1, the left of v3, 
and between v2 and v3 (see Fig.~S8(g)).

The behaviors are compared with those of a conventional particle model,
which determines the motion of the focal agent by averaging the forces 
exerted by all virtual agents.
For two virtual agents, we observed a bifurcation
as shown in Fig.~\ref{bifurcation}(e), 
but the distribution $P(x;L)$ is determined by the equilibrium position 
of the averaged forces.
In fact, the focal agent in the conventional particle model
stays around a certain position between virtual agents
instead of showing a back-and-forth motion 
 (see Fig.~S3(c),(f)).
For three virtual agents, $P(x;L)$
shows a two-way fork pattern that closes at large $L$
as shown in Fig.~\ref{bifurcation}(f),
instead of a three-way fork.
The focal agent does not select and approach the left and right virtual agents.

\subsection{Collective motion}

Next we study collective motion of many agents.
For the initial condition, we randomly positioned 100 agents
in a circle with a radius of 7 length unit.
They have the same initial speed $v_i=v_0$ 
with the moving directions $\theta_i$
and angle of visual attention $\phi_i$
uniformly distributed in $[-\pi,\pi]$.

Varying the parameters $(\chi,\omega_o)$,
we obtained the collective patterns shown 
in Figs.~\ref{snap pattern}(a),S9.
They are classified into
(i)~vortex: agents rotating around a common axis,
(ii) polarized school: agents aligned and moving coherently in the same direction,
(iii) swarm: randomly oriented agents,
(iv) turning: an elongated and curved cluster 
that intermittently develops from a polarized school
(see also Fig.~S9(a)),
and
(v) unsteady aggregation: 
a vortex collapses and then is 
transformed into a polarized school,  which becomes a vortex again.  
This cycle occurs repeatedly with irregular time intervals 
(see Fig.~S9(b)).
See Supplemental Movies. S1-S5 for the dynamics of (i)-(v), 
respectively.
The size of these patterns read from Fig.~\ref{snap pattern} is on the order of 10 length unit.
The parameters $\chi=0.3,\omega_o=1.0$ 
that were used to study selective decision making in the previous section
corresponds to the vortex pattern. 

As quantitative measures of the patterns, 
we introduce the polar order parameter $P$ 
and the rotational order parameter $M$
(see \textit{Materials and Methods}).
For the steady patterns (i)-(iii),
the order parameters rapidly converge to constants 
(see Fig.~S10).
For (iv), both $P(t)$ and $M(t)$ exhibit spikes 
that correspond to the emergence of curved clusters from a polarized school.
For (v), both $P(t)$ and $M(t)$  oscillate with large amplitudes. 
For each parameter set ($\chi,\omega_o$),
we performed 25 simulations 
in the time domain $t\in[0,800]$ to 
obtain the average values of $P$ and $M$. 
The plot of $P$ and $M$ 
in the ranges
$\chi\in[0,1]$, $\omega_o\in[0,3]$ 
is shown in Fig.~\ref{snap pattern}(b).
We consider the noiseless case ($D_v=D_\theta=0$) 
to maximize the stability of the clusters.
The plot gives a phase diagram of the collective patterns.
As the strength of alignment $\omega_o$ increases, 
the agents get aligned and 
the cluster changes from a swarm to a vortex, 
and then to a polarized school or turning.
For $\chi\lesssim0.5$, 
the vortex emerges for $\omega_o\lesssim1.5$-2.0, 
and
the polarized school for $\omega_o\gtrsim1.5$-2.0.
For $\chi\gtrsim0.5$, we obtain unsteady aggregation 
for $1.5\lesssim\omega_o\lesssim2.0$,
and turning for $\omega_o\gtrsim2.0$.
Note that the emergence of swarm is controlled by $\omega_o$
and is almost independent of $\chi$.
See also Fig.~S11 for a phase diagram based on
other quantities.

The dependence on the number of agents $N$ is studied by 
keeping the same number density in the initial condition.
As shown in Fig.~S15(a),(b),
the cluster tends to be stable for $50\lesssim N\lesssim200$,
but splits frequently for both $4\lesssim N\lesssim50$ and $N\gtrsim200$.
(See \textit{SI Text} and Figs.~S12,S14 
for a detailed analysis of the splitting and 
its dependence on the other parameters.)

\subsection{Visual information and topological distance}
\begin{figure}[!t]
\centering
\includegraphics[width=\linewidth]{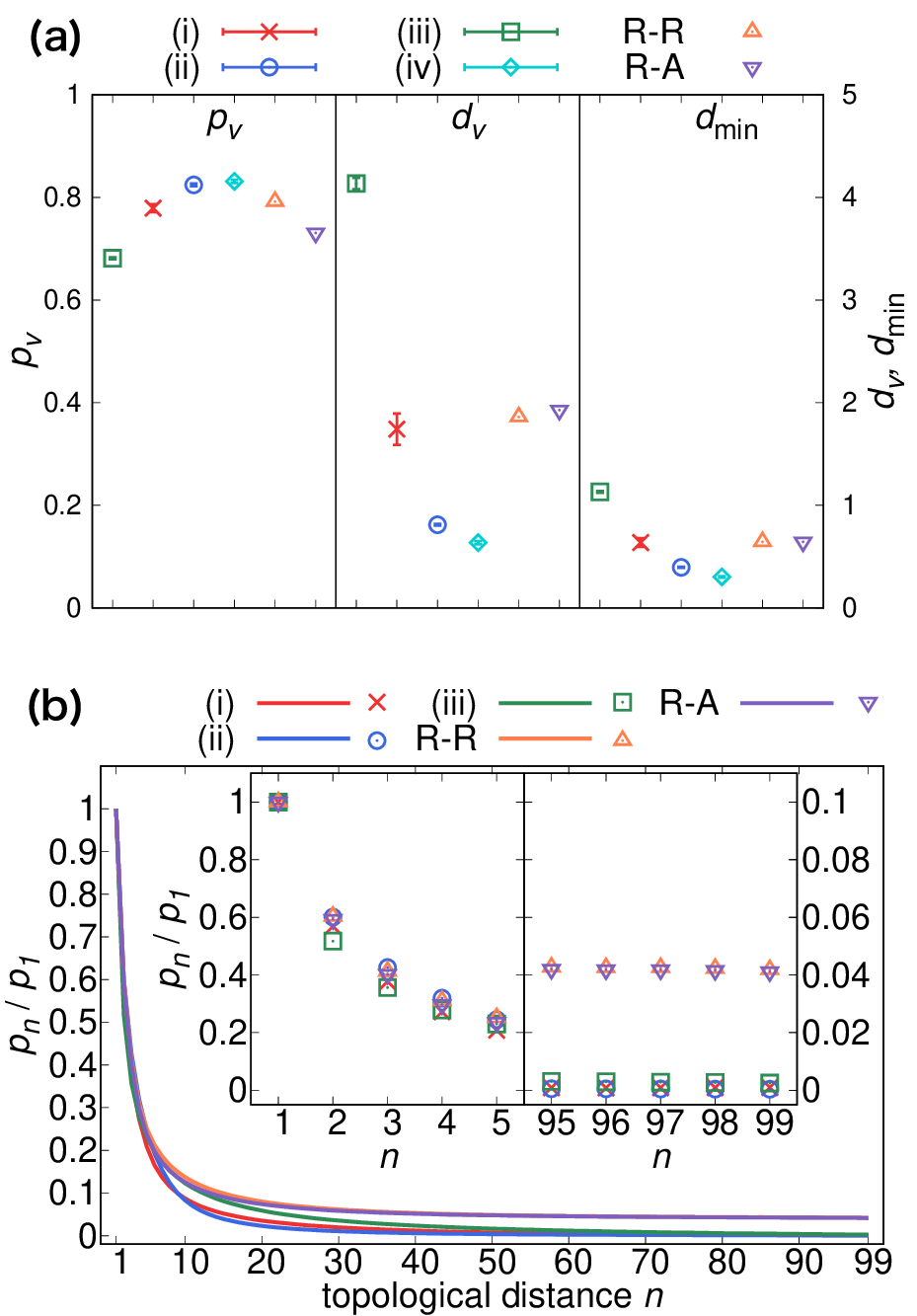}
\caption{
Visual information of an agent in a 
(i) vortex (red) (ii) polarized school (blue), (iii) swarm (green), and
(iv) turning cluster (cyan).
R-R (orange) and R-A (purple) are the abbreviations of the random-random case and the random-aligned case (see text).
(a) The occupancy ratio $p_v$,
the average distance $d_v$ and the minimum distance $d_{\mathrm{min}}$ to the neighbors.
The error bar represents the standard deviation.
(b) The relative occupancy ratio $p_n$ normalized by $p_1$.
Insets show the enlarged figure for the topological distance
$n\in[0,5]$ and $n\in[95,99]$.
}
\label{visual info}
\end{figure}

Here, we analyze the visual information of an agent 
in a cluster with 100 agents.
We measure the occupancy ratio $p_v$ as the angular fraction of the images in
the field of view,
the average distance 
from the eye to the agents 
$d_v$ 
estimated by the vertical angular diameter,
and the minimum distance to the neighbors $d_{\mathrm{min}}$.
As shown in Fig.~\ref{visual info}(a),
$p_v$ is smaller 
and $d_v$ and $d_{\mathrm{min}}$ are larger
in the order of 
(iii) swarm, (i) vortex, (ii) polarized school, and (iv) turning.
In particular, $d_v$ is close to the equilibrium distances of the forces $r_e=2$ and $\rho_e=1$
for (i) and (ii),(iv), respectively.

To study the visual screening effect by the neighbors,
we define the relative occupancy ratio $p_n$ of 
the $n$-th nearest neighbor ($n$NN) 
as
the fraction of the number of bins occupied by the image of $n$NN 
in the bins occupied by all neighbors.
We call $n$ the topological distance. 
As shown in Fig.~\ref{visual info}(b),
$p_n$ decays to almost zero at
$n=99$ for the patterns (i),(ii), and (iii),
which means that an agent cannot see all the neighbors 
in the cluster at a time.
For comparison,  we also study the case where 
100 virtual agents are randomly positioned
in a circle with a radius which is 7 length unit.
Their orientation is either random (referred to as the random-random case)
or aligned (the random-aligned case).
Interestingly, for both cases, $p_n$ reaches to a
non-vanishing constant at $n=99$ (see Fig.~\ref{visual info}(b)),
even though $p_v$ is larger and $d_v,d_{\mathrm{min}}$ are smaller than
that of a swarm (see Fig.~\ref{visual info}(a)).
It indicates that the visual field of an agent is screened more strongly
by the clustering of real agents. 

Next we verify the role of the visual attention in the screening effect. 
The probability for the angle of visual attention 
to lie within the bins occupied by $n$NN
also reaches zero as $n$ is increased (see Fig.~S16(d)).
Furthermore, the speeding force as a function of the topological distance $n$
(Fig.~S17) is a strong attraction for large $n$. 
However, because the probability for the visual attention is small for large $n$,
the expected value of the force is dominated by
the repulsive force from near neighbors, which occupy most of the visual field.

\subsection{Force map for pairwise and three-body interaction}

\begin{figure}[!t]
\centering
\includegraphics[width=\linewidth]{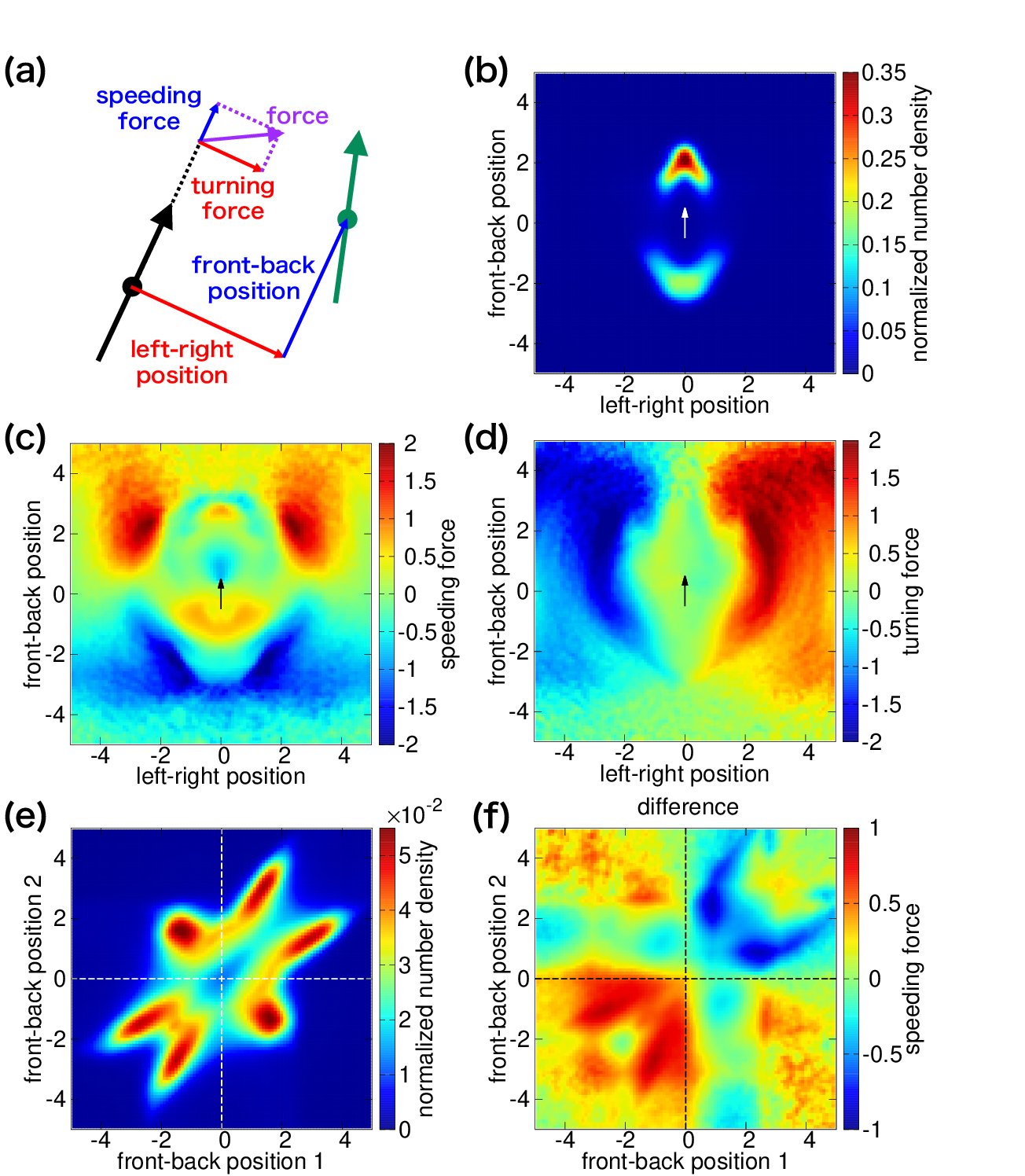}
\caption{
(a)
Definition of the force acting on the
the focal agent (black) 
and relative position of the neighbor (green).
For two-body interaction,
(b) the positional distribution, 
(c) speeding force,
(d) turning force
are mapped with respect to the relative position.
The white or the black arrow in the center of each map 
represents the focal agent.
For three-body interaction,
(e) the positional distribution
and
(f) the difference speeding force
are mapped with respect to the front-back positions of 
neighbor 1 and neighbor 2.
The parameters are $\chi=0.3,\omega_o=1.0$.
}
\label{force map}
\end{figure}

Finally, in the presence of one or two neighbor agents,
we consider their positional distribution (number density) 
and forces exerted by them. They are mapped
as functions of the front-back position and the left-right position of the neighbors.
Following the experiment~\cite{Katz2011},
we measure the force by the acceleration of the focal agent 
and divide it into the speeding force and turning force,
which are the components parallel and perpendicular to the direction of motion, 
respectively (see Fig.~\ref{force map}(a)).

For two agents, 
the positional distribution of the neighbor 
has two peaks in the front and rear.
The front peak is higher than the rear one, as shown in Fig.~\ref{force map}(b).
Regarding the force map, the speeding force shows strong attraction
at the left- and right-front and at the center-rear  (see Fig.~\ref{force map}(c)).
This forward-backward asymmetry is originated from
the self-propelled force.
As shown in Fig.~S18(a), if we subtract the self-propelled force,
the force map becomes forward-backward symmetric
by construction of the speeding force of our model (see Eq.~(\ref{eq F})).
The turning force is strong at the left- and right-front  (see Fig.~\ref{force map}(d)).
This is interpreted by the alignment term,
which is strong at the frontal side (see Eq.~(\ref{eq Omgo})).
Moreover, as shown in Fig.~S18(b),(d),
the attractive speeding force and the attractive turning force
increase with the relative speed of the neighbor,
although we did not explicitly introduce such a dependence 
in the model equations~(\ref{eq v}),(\ref{eq tht}).
(See \textit{SI Text} and Figs.~S18(c),(e)
for the relative heading dependence of a neighbor.)

Next, we focus on three-body interaction.
In particular, 
we analyze the relation between 
the front-back positions of the two neighbors 
and three-body interaction.
As shown in Fig.~\ref{force map}(e),
the normalized number density exhibits a characteristic aster pattern:
this pattern indicates 
that the three agents tend to make a line-formation 
with a certain front-back distance between each pair.
The speeding force of three-body interaction nearly vanishes
at the equilibrium position which corresponds to the peak of the normalized number density
(see Fig.~S19(d)).
To clarify the nature of three-body interaction,
we subtract the averaged pairwise speeding forces from the total force.
The pairwise forces are calculated assuming that only one neighbor exists at a time
(see Fig.~S19(e)).
As shown in Fig.~\ref{force map}(f),
the difference force reinforces the repulsion
when both neighbors are at the front or rear of an agent,
and also generates a restitution force when the two neighbors are in the front and rear.
Furthermore, the turning component of the difference force is relatively small 
compared to the speeding component
(see Fig.~S19(i)).

\section{Discussion}
Our agent-based model incorporates selective
decision making and collective motion.
The visual signal
is inspired by the behaviors of ganglion cells and eye system.
In particular,
we introduced the motion of the direction of visual attention.
The interaction forces are
modeled according to the experimental knowledge,
and are averaged over the resolution region of the visual attention,
instead of the conventional integrated pairwise interaction over all neighbors.

To confirm that our model actually shows the selective decision making,
we used the virtual agents method.
A focal agent selects a virtual agent spontaneously with the dynamical back-and-forth motion
(see Figs.~S3(a),(d)), and
the marginal distributions exhibit the bifurcation behavior (see Fig.~\ref{bifurcation}(c)-(d))
similar to those observed in fish and insects~\cite{Sridhar2021}.
In the case of zebrafish, the two- and three-way bifurcation 
for two and three virtual fish, respectively, 
occurs at $L\sim6$~BL~\cite{Sridhar2021}.
On the other hand, the bifurcation occurs at $L\sim3$~BL in our model.
The difference might be explained by the resolution of the eyes;
fish with a bigger body  (e.g. golden shiner) has better resolution than zebrafish~\cite{Pita2015},
for which our model might show a better agreement.
In the presence of
three virtual agents, the probability near the center one becomes small for large $L$,
in agreement with the zebrafish experiment~\cite{Oscar2023}.
For an asymmetrical configuration of virtual agents, 
the probability distribution shows three peaks (see Fig.~S8(g)),
which also catches the tendency 
for zebrafish~\cite{Sridhar2021}.
Furthermore, we have shown clearly that the conventional model 
which simply averages pairwise interactions from neighbor fish
cannot reproduce the bifurcation process (see Fig.~\ref{bifurcation}(e)-(f)).

Our model shows various patterns of collective motion 
(vortex, polarized school, swarm, and turning)
as shown in Fig.~\ref{snap pattern}(a).
The size of the vortex and polarized school with 100 agents is about $\sim10$ BL, 
in agreement with the size of collective 
patterns of 70 and 150 golden shiners
(read from the Figures and Movies in Ref.~\cite{Tunstrom2013}).
In the turning pattern, the cluster rapidly turns (see Fig.~S9(a)),
and it is different from the turning phase found in the previous model~\cite{Filella2018}
that maintains the curved shape.
The parameter  
$\chi$ that characterizes the anisotropy of the ganglion cell density
is estimated as about $1/3$ from an experiment
(see Table~\ref{experimentalvalue} and \textit{SI Text}),
and the vortex, polarized school, and swarm emerge in this $\chi$ region (see Fig.~\ref{snap pattern}(b)).
Regarding the dependence on the number of agents $N$,
the cluster is stable for
 $50\lesssim N\lesssim200$, 
 while a larger cluster tends to split into small clusters
 (see Fig.~S15).
The instability might be caused by
confinement in 2D~\cite{Calovi2014,Filella2018};
{in  fact, a 3D model can reproduce a stable vortex for up to 10000 agents}~\cite{Ito2022a,Ito2022b}.
For $N\lesssim50$,
we obtained frequent splitting,
but it might be suppressed by
the noise-governed interaction method 
as hypothesized for
cichlid (\textit{Etroplus suratensis})~\cite{Jhawar2020}.

The ratio of the visual field which is not filled by the other agents
is $1-p_v\sim0.2$ (see Fig.~\ref{visual info}(a)),
which is consistent
with the ratio $\sim0.2$-0.4 for 70-151 golden shiners~\cite{Davidson2021}.
In a cluster, the visual field of an agent is screened more by the neighbors
comparing with the random configuration (see Fig.~\ref{visual info}(b)).
Regarding the topological distance,
the first nearest neighbor contributes most to
the repulsive speeding force and screening of the attraction
(see Fig.~S17(c)).
On the other hand, 
for the angular velocity
is not determined solely by the nearest neighbor,
but several neighbors contribute nearly equally.
(see Fig.~S17(d)).
In fact, the first nearest neighbor dominates the interaction in the case of mosquitofish~\cite{Herbert2011},
and this dominance might be also  
responsible for the strong repulsive speeding force 
compared with the attractive speeding force and the turning force
in a school of golden shiner~\cite{Katz2011}.

Next we compare 
the positional distribution of neighbor fish and the force map,
with  the experimental results on golden shiner~\cite{Katz2011}.
For the positional distribution, our results reproduce
the two peaks in the front and rear 
including their distances $\sim 2$ BL.
We also reproduced the aster pattern in the correlation map of the
front-back position of two neighbors (see Fig.~\ref{force map}(e)).
In our model, the neighbor density is high in the front
for both two and three agents
(see Figs.~\ref{force map}(b),S19(a)),
but golden shiner shows a forward-backward symmetrical distribution.
Also, the correlation map of the left-right positions of two neighbors
shows a peak at the center in our model (Fig.~S19(f)), but 
the peak is split in the experiment.

We plotted the force map by the same method used in the experiment.
For the speeding force, our model reproduces the small peaks 
in the front and rear due to the repulsion and
broad peaks in the far region due to the attractive force.
However,  the experimental force map has 
a front-back symmetry of the speeding force,
while our model shows asymmetrical patterns.
The symmetry might be reproduced by incorporating anisotropic interaction 
in the present model.
For dependence on the relative speed of a neighbor,
the attraction increases with the relative speed
 (see Fig.~S18(b),(d)),
which is consistent with
the experimental results.
For three-body interaction,
we reproduced the restitution
speeding force (see Fig.~\ref{force map}(f))
and the fact that the turning force is given by
the averaged pairwise force
 (see Fig.~S19(i)).
However, the reinforcement of the repulsive speeding force
by the front or rear neighbors
(see Fig.~\ref{force map}(f)) 
is small or vanishing in the experiment. 

In summary, our visual model
comprehensively
reproduces the cooperative behaviors of fish,
and give insight on decision making,  collective motion,
topological nature of interaction, and 
positional and force distributions.
Our results are mostly consistent
with the experiments, 
but additional elements in the interaction would be necessary 
to resolve the remaining issues.
Inclusion of
(i) complex neural system for integrating  information 
from the left and right eyes~\cite{Harpaz2021b},
(ii) more precise 
dependences on 
the angular position 
and relative heading of a neighbor~\cite{Calovi2018},
(iii) hydrodynamic signals via
the lateral line~\cite{Partridge1980},
and (iv) direct hydrodynamic interaction 
by the reverse Kármán vortex~\cite{Li2020,Ito2023}
will be  interesting future directions.

\section{Materials and Methods}
Here, we show the details of our model and the quantities measured in the simulation.
See \textit{SI Text} for more details of
the definition of formulas, parameters, and the other information.

\subsection{the visual signal}
The ganglion cell density is represented by $D$, which is a periodic function of $\phi_\mu$:
\begin{equation}
\label{eq D}
D(\phi_{\mu};\chi)=\frac{1+\chi\cos\phi_\mu}{1+\chi},
\end{equation}
where $\chi\in[0,1]$ is the anisotropy parameter.
The ratio between the cell densities at the front and rear side of the retina
is given by
$D(\pm\pi;\chi) =(1-\chi)/(1+\chi)$.

Dependence on the vertical angular diameter
is given by the function
\begin{equation}
\label{eq A}
A(\delta_{i,\mu}^\perp)=\hat{A}\frac{r_0}{r_0+r_{i,\mu}},
\end{equation}
where
\begin{equation}
r_{i,\mu} =\frac{h_b}{2\tan(\delta_{i,\mu}^\perp/2)}
\end{equation}
is 
the relative distance of a neighbor,
and
$r_0$ is the limiting distance at which a body length can be identified.

Dependence on the relative speed of a neighbor
 is described by
\begin{equation}
\label{eq U}
U(\delta_{i,\mu}^\perp,u_{i,\mu};\beta)=(1-e^{-r_{i,\mu}/r_a})+e^{-r_{i,\mu}/r_a}e^{\beta\{(u_{i,\mu}/v_0)-1\}}
\end{equation}
is an increasing function of $u_{i,\mu}$ controlled by $\beta\geq0$.
Here, $r_a$ corresponds to the distance
at which the pattern on the body of neighbor can be detected and the speed can be measured.

\subsection{the equation of motion}

The speeding force in Eq.~(\ref{eq v}) is given by
\begin{equation}
\label{eq F}
F(\phi_\mu,\delta^\perp_{i,\mu})=f(r_{i,\mu})\cos(\phi_\mu),
\end{equation}
\begin{align}
\label{eq f}
f(r)= \left\{ \begin{array}{ll}
-f_r\frac{r_e-r}{r_e} & [0<r<r_e], \\
f_a\frac{r-r_e}{r_a-r_e} & [r_e<r<r_a], \\
f_a\frac{r_a}{r} & [r_a<r],
\end{array} \right.
\end{align}
where $r_e$ is the equilibrium distance and  $r_a$ is the distance
that gives
 the maximum attractive force $f_a$.
When $r_{i,\mu}=0$, $f$ gives the maximum repulsive force $-f_r$.

The angular velocity $\Omega$ in Eq.~(\ref{eq tht})
consists of the repulsion-attraction term $\Omega_{r,a}$
and the alignment term $\Omega_o$~\cite{Calovi2018}, as
\begin{equation}
\label{eq Omg}
\Omega(\phi_\mu,\delta^\perp_{i,\mu},\psi_{i,\mu})=\Omega_{r,a}(\phi_\mu,\delta^\perp_{i,\mu},\psi_{i,\mu})+\Omega_o(\phi_\mu,\delta^\perp_{i,\mu},\psi_{i,\mu}).
\end{equation}
The repulsion-attraction term is
\begin{eqnarray}
\label{eq Omgra}
\Omega_{r,a}(\phi_\mu,\delta^\perp_{i,\mu},\psi_{i,\mu})&=&\omega(r_{i,\mu})\sin\phi_\mu\\\nonumber
&&\times\left(1-e^{-r_{i,\mu}/r_a}\frac{1+\cos\psi_{i,\mu}}{2}\right),
\end{eqnarray}
\begin{align}
\label{eq omg}
\omega(r)= \left\{ \begin{array}{ll}
-\omega_r\frac{\rho_e-r}{r_e} & [0<r<\rho_e], \\
\omega_a\frac{r-\rho_e}{r_a-\rho_e} & [\rho_e<r<r_a], \\
\omega_a\frac{r_a}{r} & [r_a<r],
\end{array} \right.
\end{align}
where $\rho_e$ is the equilibrium  left-right distance, 
which is smaller than $r_e$ due to the slender body of fish.
We assume that fish can precisely detect 
the relative heading as well as the relative speed
 of a neighbor within $r_a$. 
The alignment term
is 
\begin{eqnarray}
\label{eq Omgo}
\Omega_{o}(\phi_\mu,\delta^\perp_{i,\mu},\psi_{i,\mu})&=&\omega_{o}\exp\left\{-\frac{(r_{i,\mu}-r_o)^2}{2l_o^2}\right\}\\\nonumber
&&\times\frac{1+\cos\phi_\mu}{2}\sin\psi_{i,\mu},
\end{eqnarray}
where $\rho_e<r_o<r_a$, $l_o$ is the characteristic length, and $\omega_o$ is the strength of alignment.

Finally, $\langle Q(q_{i,\mu})\rangle_\mu=\sum_{\mu\in\mathcal{R}_i}Q(q_{i,\mu})/\abs{\mathcal{R}_i}$ is the average of
a quantity $Q(q_{i,\mu})$ in the resolution angle region centered on $\phi_i$:
a set of bins $\mathcal{R}_i$ includes the bin in $[\phi_i-\delta^\parallel_0/2,\phi_i+\delta^\parallel_0/2]$,
and $\abs{\mathcal{R}_i}$ is its size.
The horizontal resolution angle 
$\delta^\parallel_0$ is related to $r_0$ as $\delta^\parallel_0=2\tan[-1](\frac{l_b}{2r_0})$.

\subsection{the order parameters}
The polar order parameter is defined as
\begin{equation}
\label{eq P}
P(t)=\frac{1}{N}\abs{\sum_{i=1}^N\bm{e}_i(t)}.
\end{equation}
It takes the maximal value 1 
when the 
agents are completely aligned,
and vanishes
when the orientation of the agents is completely random.
The rotational order parameter is defined as
\begin{equation}
\label{eq M}
M(t)=\frac{1}{N}\abs{\sum_{i=1}^N\hat{\bm{c}}_i(t)\times\bm{e}_i(t)},
\end{equation}
where $\hat{\bm{c}}_i(t)=(\bm{r}_i(t)-\bm{r}_G(t))/\abs{\bm{r}_i(t)-\bm{r}_G(t)}$ 
is the unit vector
that gives the direction 
from the center of mass position $\bm{r}_G=\sum_{i=1}^N\bm{r}_i/N$.
The rotational order parameter
is large when the agents are rotating
around a common axis in the same direction.

\begin{widetext}
\section*{Supplementary Information}
Supplementary Information is available from the below URL.
\vskip\baselineskip

\url{https://drive.google.com/drive/folders/1vaXJgxHfkGwbqdp-fourrdlSb2thIKmM?usp=share_link}

\vskip\baselineskip

\section*{acknowledgement}
We acknowledge financial support by JSPS KAKENHI Grant No. 23KJ0171 to S.I.
and support by a research environment of Tohoku University,
Division for Interdisciplinary Advanced Research and Education to S.I.

\end{widetext}



\end{document}